# A Suggestive Way of Deriving the Quantum Probability Rule

Roderick I. Sutherland

Centre for Time, University of Sydney, NSW 2006 Australia

rod.sutherland@sydney.edu.au

The familiar "modulus squared" form of all quantum mechanical probabilities is derived from an assumption of equal a priori probabilities concerning the final states available.

## 1. Introduction

The probability distributions of quantum theory have the general form of the square of the modulus of a wavefunction or amplitude. In basic texts on quantum mechanics, this form is simply introduced as a separate postulate once the standard description of quantum mechanical states in terms of wavefunctions or state vectors has been presented [1]. In statistical mechanics, on the other hand, the form of any probability distribution is **derived** from an assumption of equal a priori probabilities for the available states of the system in question [2]. As will be shown here, it is, in fact, possible to derive the form of the quantum probability distributions from a similar assumption. Our way of achieving this, however, involves adopting an unconventional viewpoint. Specifically, it involves giving the modulus and phase of the state vector somewhat greater significance than at present. It is therefore a matter of personal taste whether one takes this extra significance seriously or prefers to continue treating the quantum probability rule as a separate postulate. Either way, though, the proposed derivation highlights a previously ignored possibility inherent in the structure of quantum theory.

## 2. Some further preliminary comments

For simplicity, the argument will take the following specific form. Firstly, it will be formulated in terms of a measurement performed on a single particle, rather than on a more complicated system. (From the form of the argument, however, it will be clear that it also holds in the more general case without significant alteration.) Secondly, the observable discussed will be taken to have a discrete spectrum of eigenvalues. (Generalization to the continuous case also presents no difficulties.) Thirdly, the argument will be formulated in terms of the state of the observed particle, with no reference to the state of the measuring apparatus. (It can, however, be equally well formulated in terms of the overall state of the particle and apparatus together, as will be outlined in the final discussion.) A further point to note is that the derivation is independent of whether the results of measurements arise spontaneously for no reason whatsoever, as is normally assumed, or whether they arise from some underlying deterministic mechanism.

## 3. Postulates

The postulates on which our derivation is based will now be formulated. Consider an ensemble of particles all described by the initial state $|\psi\rangle$. One normally takes the view that a state vector such as $|\psi\rangle$ is not affected by multiplication by a constant complex factor $re^{i\theta}$, since (i) an arbitrary phase factor $e^{i\theta}$ has no effect on observable probabilities and (ii) any inappropriate modulus factor r is eliminated by the process of normalizing the state vector to obtain sensible probabilities. Nevertheless, it will be useful here to pursue a rather unorthodox viewpoint instead. This entails proposing that the absolute modulus and phase of a particle's state vector are meaningful and that they give rise to the familiar form of the quantum probability expressions. In accordance with this view, explicit modulus and phase values will be introduced by taking the actual state of a particle to be the standard state vector (this being normalized) multiplied by an appropriate complex factor. Thus we have the following:

**Postulate 1:**

The initial state of each particle in our ensemble is actually of the form:

$$|S_0\rangle = r_0 \exp(i\theta_0) |\psi\rangle$$

where $|\psi\rangle$ is the usual state vector and $r_0$ and $\theta_0$ are unknown real constants which may vary from one particle to another.

Returning to the standard formalism, $|\psi\rangle$ can be expressed in terms of the set of eigenstates $|j\rangle$ of the observable to be measured, as follows:

$$|\psi\rangle = \sum_j |j\rangle\langle j|\psi\rangle \qquad (1)$$

(It is assumed here that the eigenstates form an orthonormal set.) The coefficients $\langle j|\psi\rangle$ of this series expansion then yield the probability of each outcome via

$$P(j) = |\langle j|\psi\rangle|^2$$

(this last expression being what we intend to derive). Now, suppose that the result of the measurement is the $n^{th}$ eigenvalue. One normally then takes the resulting state vector to be $|n\rangle$. In view of our postulate above, however, it is important to be precise here about the effect a measurement has. Strictly speaking, the effect of a measurement that yields the $n^{th}$ eigenvalue is to eliminate the possibility of all other eigenstates $j \neq n$ so that, in the expansion (1), only the term $|n\rangle\langle n|\psi\rangle$ remains. From this perspective, and remaining consistent with our unorthodox viewpoint, the resulting state after the measurement can be considered to be $|n\rangle\langle n|\psi\rangle$, not just $|n\rangle$. Of course, to predict the correct probabilities for a subsequent measurement, it is obviously necessary to normalize the resulting state back to $|n\rangle$ again. Nevertheless, this does not preclude us from using the absolute, unnormalized state for other purposes[1].

In the light of the above, we now state:

**Postulate 2:**

For a particle initially in the state $|S_0\rangle$ of postulate 1, the final state (immediately after the measurement) is

$$|S\rangle = r_0 \exp(i\theta_0)|n\rangle\langle n|\psi\rangle \qquad (2)$$

where n is the eigenvalue found for that particle.

Here we have simply taken the state $|S_0\rangle$ in postulate 1 and replaced $|\psi\rangle$ with $|n\rangle\langle n|\psi\rangle$.

Before stating the third and final postulate, it is convenient to rewrite the final state $|S\rangle$ more simply by introducing the following polar and cartesian notation which absorbs the factor $\langle n|\psi\rangle$ in (2):

$$|S\rangle \equiv r \exp(i\theta)|n\rangle \qquad (3)$$

$$\equiv (x + iy)|n\rangle \qquad (4)$$

(Here $r, \theta, x$ and $y$ are all real.)[2] Thus, looking at (4), the state of any particle after the measurement is completely specified by the three values x,y and n. We can now state our last postulate.

---

[1] Some measurement processes are, of course, destructive in that they immediately disrupt the eigenstate they produce. This means that neither $|n\rangle$ (in the standard theory) nor $|n\rangle\langle n|\psi\rangle$ (in the present approach) is a correct description at later times. One can, however, view any such disruption as a secondary effect occurring after the pure reduction from $|\psi\rangle$ to $|n\rangle$ or $|n\rangle\langle n|\psi\rangle$.

[2] Note from equations (2) and (3) that the modulus r and phase $\theta$ of the final state corresponding to eigenvalue n are both completely determined once the initial state's modulus $r_0$ and its phase $\theta_0$ are both specified. In particular, the final state's modulus is $r = r_0|\langle n|\psi\rangle|$ and its phase is $\theta = \theta_0 +$ phase of $\langle n|\psi\rangle$. The measurement outcome for the eigenvalue n is not determined by the initial state, but the values of r and $\theta$ are determined (by $r_0, \theta_0$ and n) once a value for n arises.

**Postulate 3:**

All possible final states (x,y,n) have equal a priori probabilities.

In other words, we assume a condition of randomness such that a particle is equally likely to finish up in any one of the final states available[3]. This may be compared with the postulate of equal a priori probabilities in statistical mechanics, which states that an isolated system in equilibrium is equally likely to be in any one of its accessible states. Our justification for postulate 3 is that all states should, in the absence of anything to break the symmetry, be treated on the same footing and therefore be considered equally probable.

**4. The derivation**

From the above postulates we will now derive the familiar form

$$P(n) = |\langle n|\psi\rangle|^2$$

of the quantum probability laws. To avoid problems with normalizability, it is convenient to impose an interim restriction on the possible range of initial states. The (hidden) modulus $r_0$ will be temporarily limited to

$$r_0 \leq R \tag{5}$$

where R is a finite cut-off value. Our final result will be found to be independent of this arbitrary cut-off. Comparing (2) and (3) we have

$$r = r_0 |\langle n|\psi\rangle|$$

which enables us to write down the corresponding restriction that (5) imposes on the **final** modulus r:

$$r \leq |\langle n|\psi\rangle| R$$

Simplifying the notation via

$$|\langle n|\psi\rangle| \equiv A \tag{6}$$

this becomes

$$r \leq AR$$

or, in terms of the cartesian variables of equation (4),

$$x^2 + y^2 \leq A^2 R^2$$

Using this last result, postulate 3 can now be written more specifically in terms of a joint probability density $P(x, y, n)$ as follows:

**Postulate 3':**

$$P(x, y, n) = \begin{cases} K & \text{for } x^2 + y^2 \leq A^2 R^2 \\ O & \text{otherwise} \end{cases}$$

where K is a normalization constant to be determined.

It is now a simple matter to integrate this distribution over all $x$ and $y$ to obtain the desired probability $P(n)$. We have

$$P(n) = \int\int_{-\infty}^{+\infty} P(x, y, n)\, dx\, dy = \iint_{x^2 + y^2 \leq A^2 R^2} K\, dx\, dy$$

and, since the integral is over a circular area of radius AR, this yields

$$P(n) = K \pi A^2 R^2 \tag{7}$$

The constant K can be evaluated by summing over all values of n (and using (6)):

$$\sum_n P(n) = \sum_n K \pi A^2 R^2 = K \pi R^2 \sum_n |\langle n | \psi \rangle|^2 = K \pi R^2$$

with the requirement that the total probability be equal to 1 then implying

$$K = (\pi R^2)^{-1}$$

Substituting this back into (7), we finally obtain

$$P(n) = A^2$$
$$= |\langle n | \psi \rangle|^2$$

Thus the familiar form of quantum mechanical probabilities has been deduced by assuming equal a priori probabilities for all measurement outcomes, in conjunction with assigning some absolute significance to the modulus and phase of the state vector. Note that this result is independent of the cut-off value R imposed temporarily on $r_0$.

---

[3] From postulate 3 it is not difficult to prove that all of the possible **initial** states also have equal probabilities, as is shown in the Appendix.

## 5. Discussion

The essence of the above line of reasoning may be visualized by considering the space defined by the three variables x,y and n. Limiting ourselves to the portion of this space left available to particles by the cut-off value, the probability of a particular measurement outcome n is determined by the relative amount of (x,y,n) space that is associated with that n value. This amount is a circular area of x,y values related to the amplitude $\langle n|\psi\rangle$. In particular, the radius of the circle is proportional to $|\langle n|\psi\rangle|$ and so the area is proportional to $|\langle n|\psi\rangle|^2$.

The derivation above has been carried out in the context of non-relativistic quantum mechanics. It can, however, be generalized without essential modification to the domains of relativity and field theory. For example, in quantum field theory we have an initial state $|\psi\rangle$ (perhaps consisting of several particles) and complex amplitudes $\langle j|\psi\rangle$ connecting $|\psi\rangle$ to the range of final states $|j\rangle$ available (perhaps consisting of various numbers of different particles). One can then associate an absolute modulus and phase $r_0 \exp(i\theta_0)$ with $|\psi\rangle$ and again derive the quantum mechanical probability rule as above.

Also, as mentioned earlier, the argument can equally well be formulated in terms of the combined particle/apparatus state. To achieve this, we associate an absolute modulus r and phase $\theta$ with both the particle's initial state $|\psi_p\rangle$ and the apparatus' initial state $|\psi_A\rangle$:

$$|S_p\rangle = r_p \exp(i\theta_p)|\psi_p\rangle$$
$$|S_A\rangle = r_A \exp(i\theta_A)|\psi_A\rangle$$

The overall initial state is then

$$|S_o\rangle = r_p \exp(i\theta_p) r_A \exp(i\theta_A)|\psi_p\rangle|\psi_A\rangle$$
$$\equiv r_0 \exp(i\theta_0)|\psi\rangle$$

and the derivation proceeds as before, with the initial state $|\psi\rangle$ and the final states $|j\rangle$ now describing the combined particle/apparatus system.

The argument presented here is thus seen to be quite general and leads us to ponder the degree of physical significance of a state vector's modulus and phase.

**Appendix**

We have assumed that all possible **final** states are equally likely. It is, of course, desirable that all the **initial** states $|S_o\rangle$ described in postulate 1 be equally likely as well. By expressing $|S_0\rangle$ in the cartesian form $(x_0 + iy_0)|\psi\rangle$, it is not difficult to verify from postulate 3 that all possible initial states $(x_0, y_0)$ do, in fact, have equal probabilities, as will be shown in this appendix.

We start by comparing equations (2) and (3) to obtain

$$r_0 \exp(i\theta_0) \langle n|\psi\rangle = r \exp(i\theta) \tag{8}$$

For convenience, we will express the complex amplitude $\langle n|\psi\rangle$ as

$$\langle n|\psi\rangle \equiv a + ib \tag{9}$$

so that equation (8) can be rewritten in the cartesian form

$$(x_0 + iy_0)(a + ib) = (x + iy)$$

Solving this for x and y yields

$$x = ax_0 - by_0$$

$$y = ay_0 + bx_0 \tag{10}$$

We are interested in obtaining an expression for the joint probability distribution $P(x_0, y_0)$. This is most easily achieved by transforming the distribution $P(x, y, n)$ of postulate 3' into the related form $P(x_0, y_0, n)$ and then summing over all n. These last two distributions are related via

$$P(x_0, y_0, n) = P(x, y, n) \frac{\partial(x, y)}{\partial(x_0, y_0)} \tag{11}$$

where the second term on the right is the Jacobian of the transformation (10). Evaluating this Jacobian we obtain

$$\frac{\partial(x, y)}{\partial(x_0, y_0)} = a^2 + b^2 = |\langle n|\psi\rangle|^2$$

where we have used equation (9). Hence, referring to postulate 3' again, equation (11) can be expressed as

$$P(x_0, y_0, n) = K |\langle n|\psi\rangle|^2$$

for values of $x_0$ and $y_0$ in the range $x_0^2 + y_0^2 \leq R^2$. Summing over all values of n then yields the final result:

$$P(x_0, y_0) = K$$

We have thus succeeded in proving that the probability of any initial state $(x_0, y_0)$ is constant, independent of the values of $x_0$ and $y_0$, which means that all initial states are equally likely.